\documentclass[conference,onecolumn]{IEEEtran}
\IEEEoverridecommandlockouts
\usepackage{cite}
\usepackage{amsmath,amssymb,amsfonts}
\usepackage{algorithmic}
\usepackage{graphicx}
\usepackage{textcomp}
\usepackage{xcolor}
\usepackage[utf8]{inputenc}
\def\BibTeX{{\rm B\kern-.05em{\sc i\kern-.025em b}\kern-.08em
    T\kern-.1667em\lower.7ex\hbox{E}\kern-.125emX}}
\DeclareMathOperator*{\argminA}{arg\,min} 

\DeclareMathOperator*{\argmaxD}{max} 

\DeclareMathOperator*{\argminD}{min} 

\DeclareMathOperator*{\mathEletter}{\mathbb{E}}
\usepackage{hyperref}

\usepackage{calrsfs}
\DeclareMathAlphabet{\pazocal}{OMS}{zplm}{m}{n}

\newcommand{\Lb}{\pazocal{L}}
\usepackage[bb=px]{mathalfa} 

\begin{document}

\title{Generation of Artificial CT Images using Patch-based Conditional Generative Adversarial Networks\\}

\author{\IEEEauthorblockN{Marija Habijan}
\IEEEauthorblockA{\textit{Faculty of Electrical Engineering, Computer } \\
\textit{Science and Information Technology Osijek}\\
\textit{J.J. Strossmayer University of Osijek}\\
Osijek, Croatia \\
marija.habijan@ferit.hr}
\and
\IEEEauthorblockN{Irena Galić}
\IEEEauthorblockA{\textit{Faculty of Electrical Engineering, Computer } \\
\textit{Science and Information Technology Osijek}\\
\textit{J.J. Strossmayer University of Osijek}\\
Osijek, Croatia \\
irena.galic@ferit.hr}
}

\maketitle

\begin{abstract}
Deep learning has a great potential to alleviate diagnosis and prognosis for various clinical procedures. However, the lack of a sufficient number of medical images is the most common obstacle in conducting image-based analysis using deep learning. Due to the annotations scarcity, semi-supervised techniques in the automatic medical analysis are getting high attention. Artificial data augmentation and generation techniques such as generative adversarial networks (GANs) may help overcome this obstacle. In this work, we present an image generation approach that uses generative adversarial networks with a conditional discriminator where segmentation masks are used as conditions for image generation. We validate the feasibility of GAN-enhanced medical image generation on whole heart computed tomography (CT) images and its seven substructures, namely: left ventricle, right ventricle, left atrium, right atrium, myocardium, pulmonary arteries, and aorta. Obtained results demonstrate the suitability of the proposed adversarial approach for the accurate generation of high-quality CT images. The presented method shows great potential to facilitate further research in the domain of artificial medical image generation. 
\end{abstract}


\section{Introduction}

Medical imaging is critical for obtaining high-quality images of nearly all visceral organs, including the brain, heart, lungs, kidneys, bones, and soft tissues. In recent years, a range of imaging modalities has employed a variety of techniques for image capture, including echocardiography, computed tomography (CT), magnetic resonance imaging (MRI), and positron emission tomography (PET). Combined with special medical software, these imaging techniques facilitate the accurate diagnosis and prognosis of different life-threatening conditions and diseases. Medical software is commonly based on two types of methods: semi-automatic and fully automatic methods. Semi-automatic methods are based on traditional image processing methods, such as active contours or different active appearance models  \cite{Kang2012HeartCA,Heimann2009StatisticalSM,Mkel2002ARO}. Their semi-automatic nature allows clinical experts to interact and manually inspect obtained results. Due to their controllability, semi-automatic methods are currently used in clinical practice.
Nevertheless, since medical images are specific to each individual patient and due to intricate tissue structures, manipulating such images using traditional image processing methods is very difficult or impossible, at least in a realistic manner. Recently, fully automatic methods based on artificial intelligence are gaining high momentum \cite{Chen2020DeepLF,Liu2021ARO,Taghanaki2020DeepSS}. Nevertheless, such methods require a large amount of medical data with corresponding annotations made by radiologists or clinical experts, which is a time-consuming process.  

The data scarcity problem presents a significant challenge for medical image processing. Recently, methods for the artificial generation of medical images have been introduced. An advantage of these approaches is that they are relatively easy to generate and control in a precise manner, which is essential for studying clinicians' cognitive and perceptual systems. However, those artificial medical images are inauthentic and entirely unlike what clinicians routinely examine. Thus, the results of these experiments fall invariably within the shadow of a doubt about their clinical applicability. Therefore, generating authentic and easily controllable medical images is critical for the entire field of medical image processing research. Alleviating constraints have only recently become more achievable, with the impressive development of Generative Adversarial Networks (GANs)  \cite{Karras2018ProgressiveGO,Karras2019ASG,Radford2016UnsupervisedRL}. 

GANs use an adversarial principle where two networks, the generator (G) and the discriminator (D), compete with each other to allow the generation of realistic-looking images with similar statistic properties compared to a sample set from random noise \cite{Mirza2014ConditionalGA}. Some examples of successful GAN applications for medical image generation and synthesis include the use of a progressive GAN for the generation of high-resolution mammography images \cite{Korkinof2018HighResolutionMS}, WGAN-gradient penalty (WGAN-GP) \cite{Middel2019SynthesisOM} that crates sharp images and limits the propensity of the model uncertainty due to the unbalanced generator and discriminator\cite{Middel2019SynthesisOM} while the experimental comparison of image synthesis using deep convolutional GANs (DCGANs), boundary equilibrium GANs (BEGANs) and Wasserstein GANs (WGANs) are given in \cite{Zhang2018MedicalIS}. In contrast, the benefits of using synthetic images as a form of data augmentation, which ultimately improves segmentation tasks, are presented in \cite{Shin2018MedicalIS,Sundaram2021GANbasedDA} while a new synthetic data augmentation method using progressive growing auxiliary classifier GAN (PG-ACGAN) is presented in \cite{Zhang2020MedicalIS}.

Motivated by previous research, in this work, we present an approach to generate artificial cardiac images using the conditional generative adversarial network. A method is proposed to generate CT images based on the segmentation masks of one or multiple cardiac structures, including myocardium (Myo), left atrium (LA), left ventricle (LV), right atrium (RA), right ventricle (RV), aorta (Ao), pulmonary arteries (PA) and whole heart (WH). 

The rest of the paper is structured in the following manner. In Section \ref{method} we present proposed method. Section \ref{implementation} gives dataset description, training and implementation details. Section \ref{results} presents obtained results. Finally, Section \ref{conclusion} gives concluding remarks and plans for future research.

\section{Method}\label{method}

\subsection{Generative Adversarial Networks}
The adversarial generative networks consist of the generator network (G) and the discriminator network (D). The role of the generator network (G) is to produce samples with data distribution highly similar to the data distribution of real, original data. In contrast, the discriminator network (D) has the task of distinguishing the synthesized samples from the original samples. The GAN loss function \cite{Radford2016UnsupervisedRL} is expressed as the discriminator binary cross entropy, as defined in Eq. \ref{equation1}.

\begin{equation}\label{equation1}
    \argminD_G \argmaxD_D \bigg[\mathEletter_{x\sim_{P_{x}}} [log D(x)] + \mathEletter_{z\sim_{P_{z}}} [log(1-D(G(z)))] \bigg]
\end{equation}

In Eq. \ref{equation1}, $P_{x}$ represents real data distribution with $x$ being a sample real image, $P_{z}$ is a random distribution and $z$ is a random latent vector. 
Further, $G(z)$ is the generator's output on input noise $z$, $D(x)$ denotes the discriminator's estimation of the probability that sample $x$ is a real image, $\mathEletter_x$ denotes expected values over all real data samples, $D(G(z))$ denotes the discriminator's estimate of the probability that a fake instance is real, and $\mathEletter_z$ is the expected value over all random inputs to the generator. Since the generator can not directly affect term $log(D(x))$, minimizing the loss is equivalent to minimizing $log(1 - D(G(z)))$. 


Similarly to GANs, the deep convolutional generative adversarial networks (DCGANs) \cite{Radford2016UnsupervisedRL} comprises of two networks, the generator (G) and the discriminator (D) which are both arbitrary \textit{deep convolutional networks} (unlike in GANs, where the networks were convolutional neural networks). 

\subsection{Conditional Generative Adversarial Networks}

\begin{figure}[h]
\includegraphics[width=.80\textwidth]{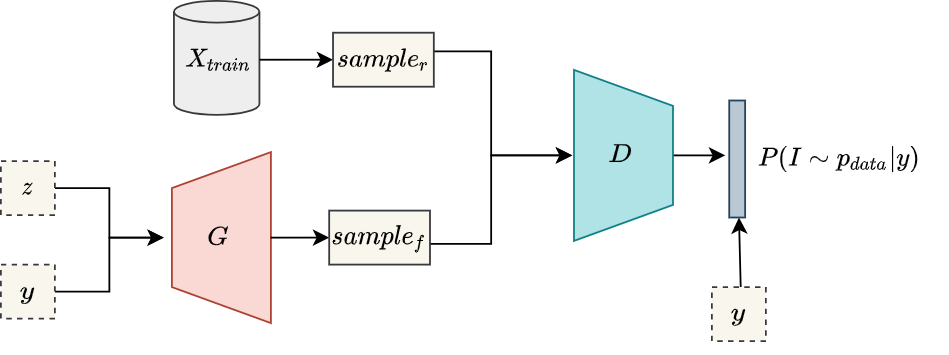}
\centering
\caption{A dense code of the output images is obtained as a combination of conditional data $y$ and noise $z$. Upsampling is performed on dense code to transform it into image space, after which, in the discriminator, images are transformed back into a dense code using convolution. Conditional data and dense code are combined in order to obtain a final prediction. The main difference between the GANs structure and cGAN is that the generator also takes conditional data as an input.}
\label{cgans}
\end{figure}

In traditional GANs, the generator's result is entirely controlled by random noise $z$. On the other hand, in conditional generative adversarial networks (cGANs), we have additional control over the generator's output with some vector $y$. In this way, the discriminator learns different values. This allows the discriminator to create a different decision boundary and forces the latter to create samples that correspond to the values of the generator's input. An illustration of cGAN is shown in Fig.~\ref{cgans}. 

Although traditional GANs generate data in a stochastic way, in this work, we choose to use a deterministic approach without randomness during training or inference. The deterministic approach allows the model to ignore any imposed randomness, which results in paired training where only exists only a single target image for each segmentation mask from our dataset \cite{Isola2017ImagetoImageTW,Abdi2019GANenhancedCE}. 

\begin{figure*}[h]
\begin{minipage}[t]{0.5\textwidth}
\includegraphics[width=\linewidth]{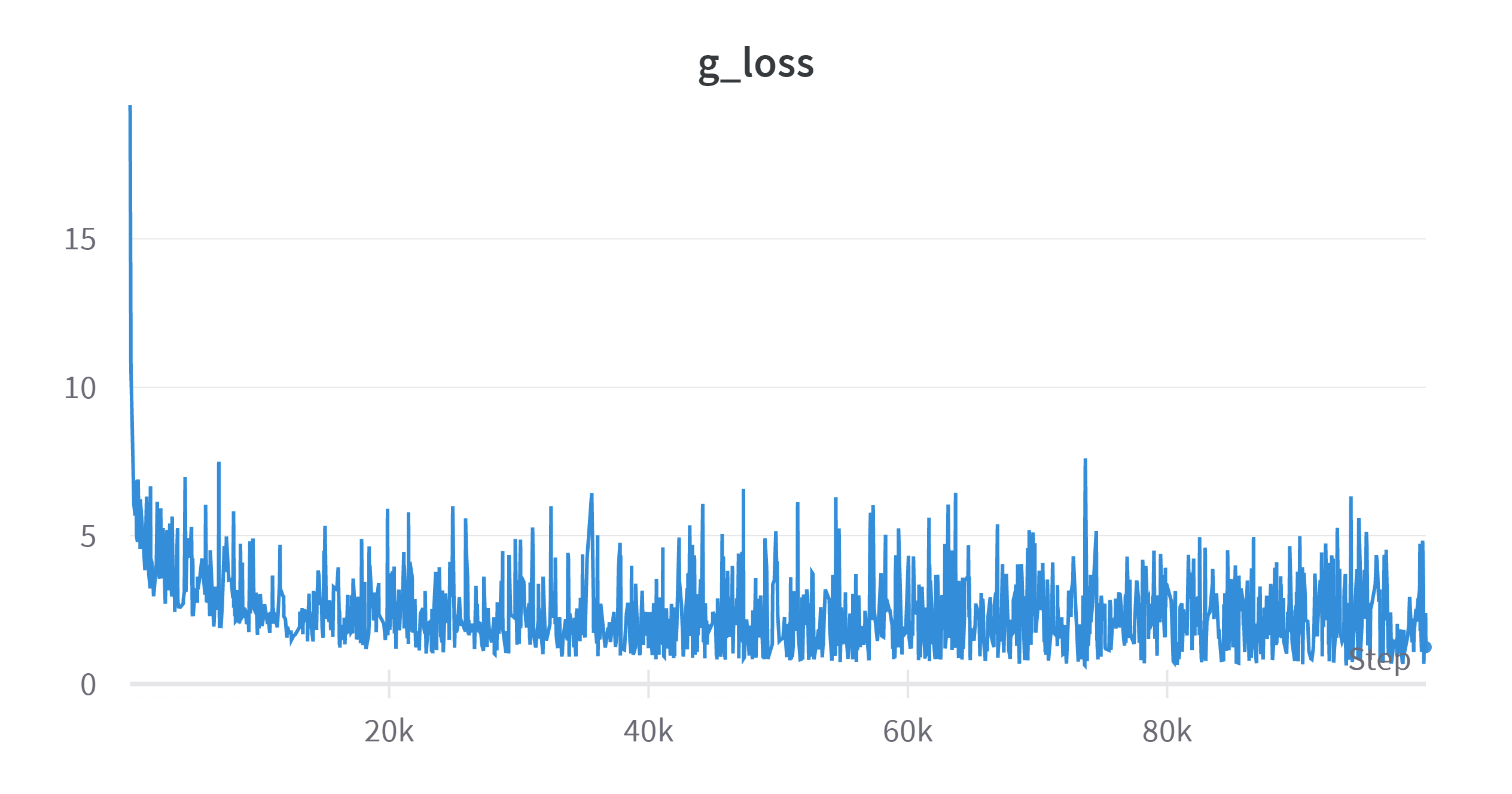}
\end{minipage}
\hfill
\begin{minipage}[t]{0.5\textwidth}
\includegraphics[width=\linewidth]{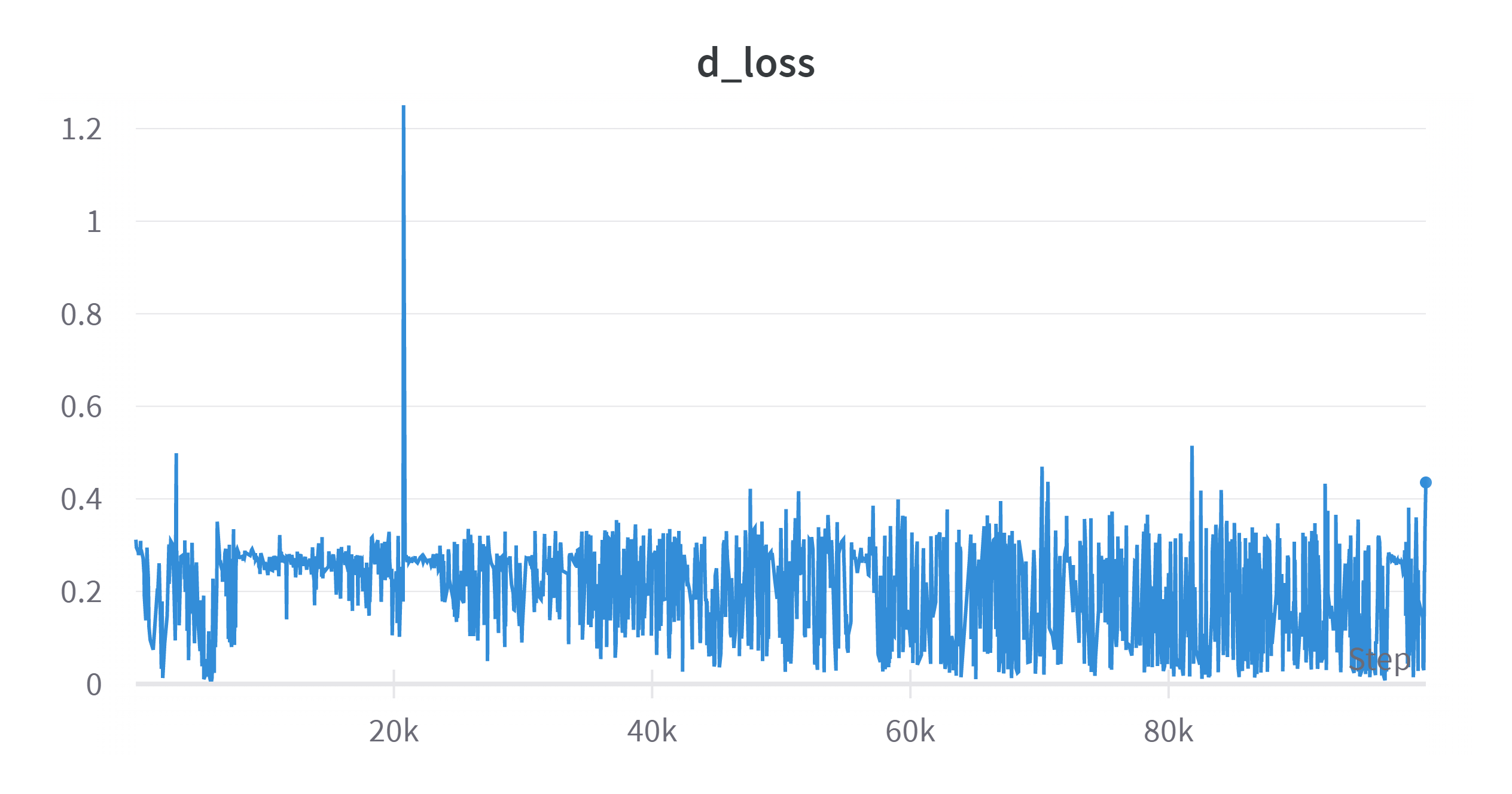}
\end{minipage}
\caption{The loss curves of the generator (left) and discriminator (right) for 100000 iterations.}\label{losses}
\end{figure*}

\begin{figure*}[h]
\begin{minipage}[t]{0.5\textwidth}
\includegraphics[width=\linewidth]{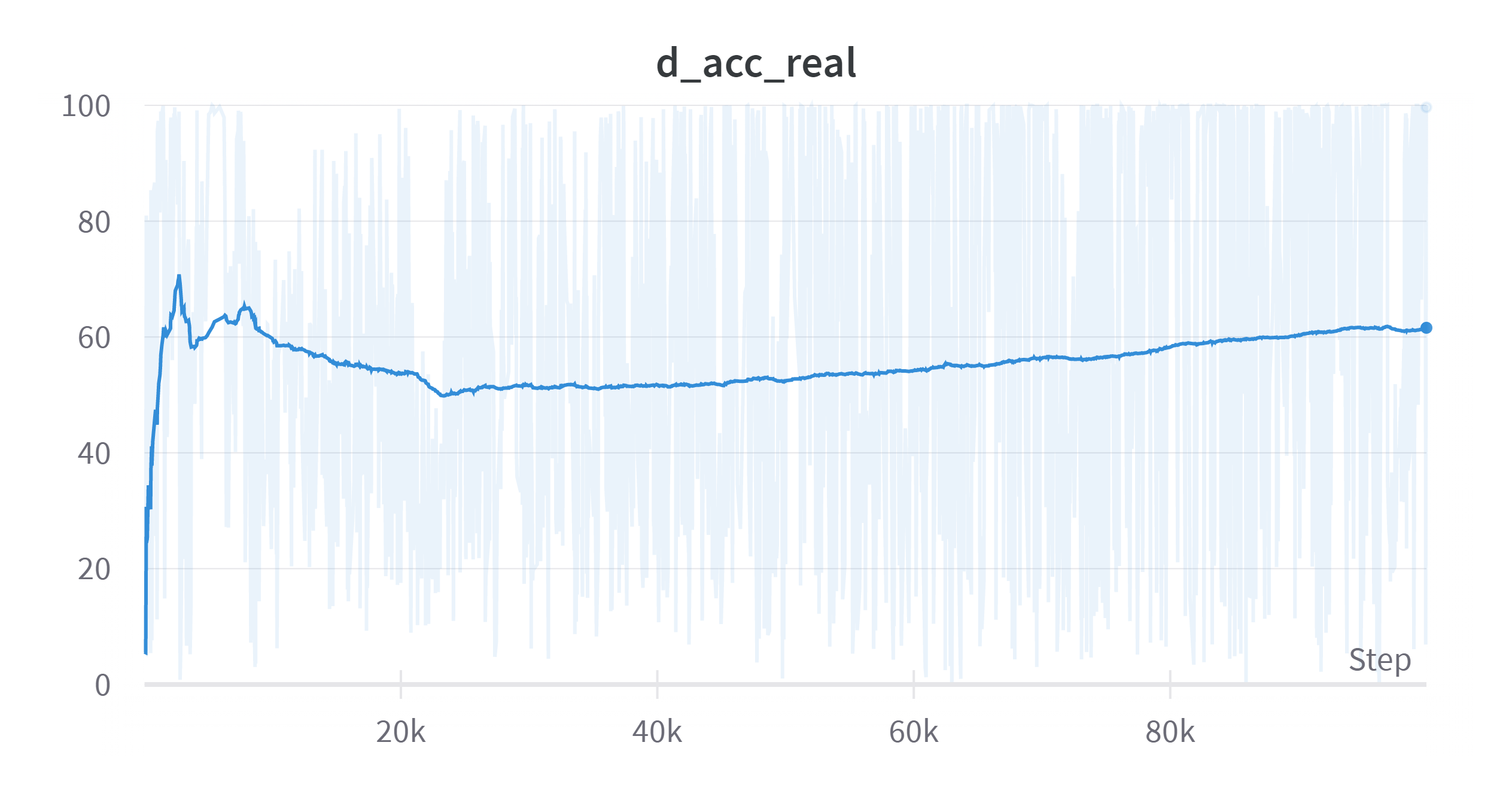}
\end{minipage}
\hfill
\begin{minipage}[t]{0.5\textwidth}
\includegraphics[width=\linewidth]{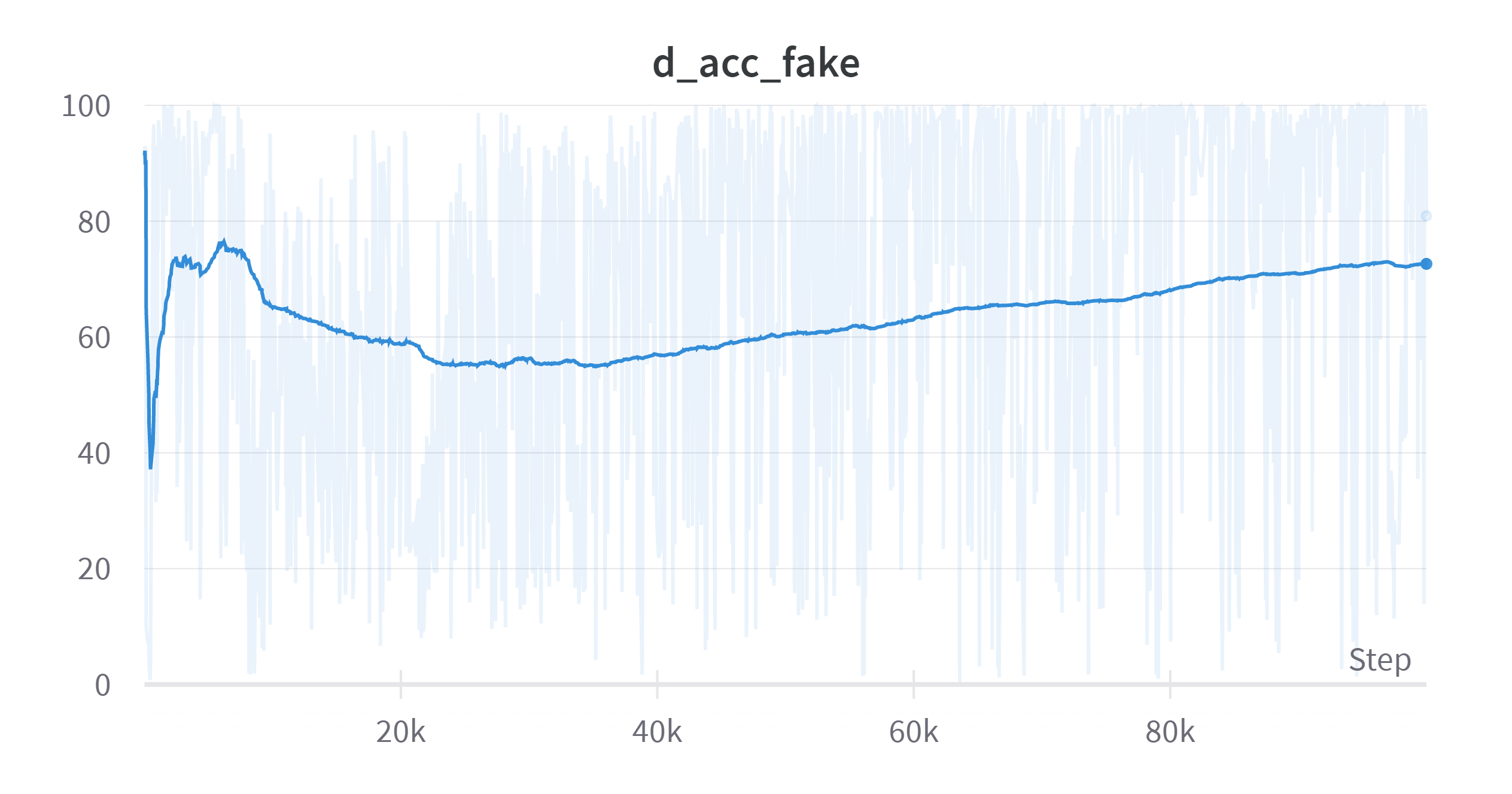}
\end{minipage}
\caption{The accuracy curves of the discriminator on the real data (left) and fake data (right) during the training evolution for 100000 iterations. The light blue represents original while single blue line is exponential moving average.  }\label{accuracy}
\end{figure*}

Furthermore, instead of the commonly used binary cross entropy (BCE) for the adversarial term, we employ the least squared error (LSE) as defined in Eq. \ref{lse1} and Eq. \ref{lse2} to force the discriminator to give specific values for the real and fake samples by pushing the generated samples closer to the real data distribution \cite{Abdi2019GANenhancedCE}.

\begin{equation}\label{lse1}
   \Lb_{cGAN}(G,D) = \mathbb{E}_{x,y} [(1-D(y,x))^{2}] + \mathbb{E}_{x} [D(y,G(y))^{2}]
\end{equation}

\begin{equation}\label{lse2}
    G* = \argminA_G \argmaxD_D \lambda  \Lb_{cGAN}(G,D) +  \Lb_{Recon}(x,G(y))
\end{equation}

Here, where $x$ and $y$ denote the target image and condition, respectively, $\lambda$ is a constant that weights the conditional adversarial term against the reconstruction term for which we use pixel-wise mean average error.

\section{Experiments and results}\label{implementation}
\subsection{Dataset Description}
In this work, we use the dataset provided by MICCAI 2017 Multi-Modality Whole Heart Segmentation Challenge \cite{dataset1}. This dataset is acquired from a real clinical environment using a cardiac CT angiography scanner. The dataset is provided and includes 20 volumetric images with an average of 350 to 500 two-dimensional slices per volume. The slices were acquired in the axial view with the pixel resolution of $512 \times 512$. The average in-plane resolution is about $0.78 \times 0.78$ mm, and the average slice thickness is $1.60$ mm. The dataset includes original CT images and their corresponding annotations of the myocardium (Myo), left atrium (LA), left ventricle (LV), right atrium (RA), right ventricle (RV), aorta (Ao), pulmonary arteries (PA) and the labels that contain the whole heart, i.e., all mentioned heart substructures together. For the preprocessing, we only use intensity normalization to the range of $[0,1]$ and resizing to $256 \times 256$. No data augmentation was used during training.

\subsection{Implementation Details}

We design our generator as a UNet-based architecture with 7 convolutional and deconvolutional layers (implemented as transposed convolutions). The discriminator works on the concatenation of the image and its condition and structurally has 5 layers. Each convolutional and deconvolutional layer (except the last layers of both the generator and the discriminator), there is a batch normalization followed by the Leaky ReLu activations. The end of both networks is a convolutional layer of stride $1$. We use Adam optimizer with the learning rate of $0.00013$, and the weight of the adversarial term was set to $\lambda = 0.012$. We used a batch size of $16$, and patches of discriminator were set to $32 \times 32$ pixels. The model was implemented with the TensorFlow and Keras deep learning library and trained on two Titan V GPUs for 100 000 iterations \footnote[1]{\url{https://github.com/mhabijan/medical\_images\_generation}}. 
Fig.~\ref{losses} and Fig.~\ref{accuracy} show the loss and accuracy curves for both generator and discriminator networks, respectively. 

\begin{figure*}[h]
\includegraphics[width=.8\textwidth]{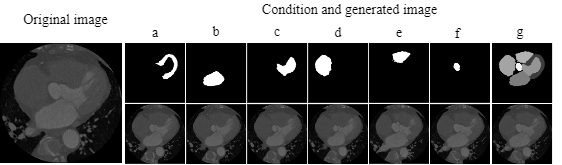}
\centering
\caption{Generated heart images, segmentation masks as condition and their corresponding original images; a) myocardium, b) left atrium, c) left ventricle, d) right atrium, e) right ventricle, f) aorta, g) whole heart.}
\label{results1}
\end{figure*}

\section{Generated Results}\label{results}

The qualitative results (visual impression) that shows comparison of original and images generated using proposed approach are shown in Fig.~\ref{results1}. As we initially assumed
the generative models have successfully and with high accuracy learned a mapping from the segmentation masks to their corresponding cardiac structures. It is important to mention that model did not see the demonstrated target images of the test set nor their conditions during training. 

\begin{figure*}[h]
\centering
\includegraphics[width=.80\textwidth]{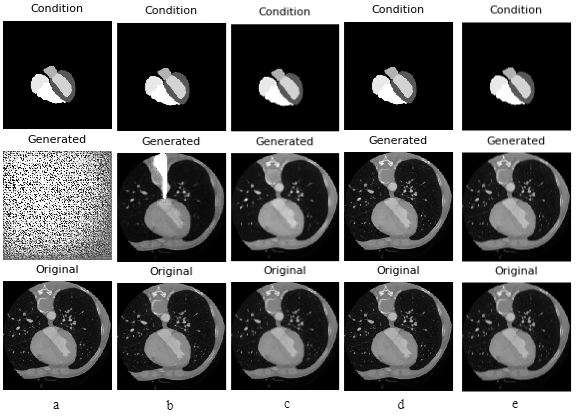}
\caption{Generated whole heart images, segmentation masks as conditions and their corresponding original images at different iterations: a) iteration 0, b) iteration 10000 , c) iteration 25000, d) iteration 50000, e) iteration 75000, f) final iteration 100000.}
\label{results_iterations}
\end{figure*}

Results of generated images during training, i.e., in different iteration steps, are shown in Fig.~\ref{results_iterations}. Here we can observe how the generation of artificial CT images develops, starting from the 0th iteration in which only a random noise was generated to the final 100,000th iteration in which the final result is achieved. It is interesting to see that already in the 10,000th iteration, a fairly clear CT image with very small artifacts is created. In the 25,000th and 50,000th iteration, small deviations appear around the left ventricle and myocardium, while in 75,000th and final 100000th iteration, there are almost no visible differences between the original and generated CT image. Furthermore, most of the generated images seem highly realistic which indicates good representation of the underlying data distribution.

\section{Conclusion} \label{conclusion}

This work presented an image generation approach that uses generative adversarial networks with a conditional discriminator where segmentation masks are used as conditions for image generation. We validate the feasibility of GAN-enhanced medical image generation on whole heart computed tomography (CT) images and its seven substructures, namely: left ventricle, right ventricle, left atrium, right atrium, myocardium, pulmonary arteries, and aorta. Obtained results demonstrate the suitability of the proposed adversarial approach for the accurate generation of high-quality CT images. 

The two main limitations of this work, low generability since we only used one dataset and modality and lack of appropriate quantitative evaluation, are good starting points for further research. Therefore, future work on this topic includes the use of a number of datasets of different organs of the human body as well as the use of different imaging modalities to prove the generality of the proposed approach. Moreover, calculation of quantitative evaluation metrics using methods like Frechet Inception Distance (FID), structural similarity (SSIM), and mean squared error are necessary to show the quality of the proposed procedure.

The offered approach demonstrates that training data can be used to generate artificial and realistic-looking medical images. We hypothesize that such technologies could be employed in the future to address privacy concerns and enable wider public data sharing. However, at the moment, there are some known issues that must be handled first.  First, legal requirements regarding data privacy differ by nation, and frequently any use of personal information.  One simple solution would be to share only produced images that have a specific degree of similarity to all original images.  Thus, reconstructing the original images would be difficult.  The remaining dilemma is what constitutes an acceptable visual difference.  A manual inspection by a physician could be one alternative.  The clinical expert could focus on removing unsuccessfully generated images as well as those that may be too similar to the source data.  Another source of concern is the anatomical accuracy of the generated data.  This, in our opinion, can only be determined by a medical expert prior to data sharing.  An expert can perform this task fairly simply for some applications, such as deciding the anatomical correctness of organ forms or limits.  Some other cases, such as tumors, may be more challenging to analyze even by an expert, and artificially created images may be too unsafe for algorithms training. Moreover, the proposed approach could alleviate further research by incorporating cross-modality images, such as mapping from CT to PET data. This may reduce the risk factor, and the cost of medical image acquisition which will provide better diagnostic decision-making. 
Finally, expansion of the proposed method to simultaneously generate original images and their corresponding annotations would significantly help to create larger data sets for training. This would ultimately be useful in developing more generalized deep learning methods that would potentially be robust enough for use in real clinical practice.

\section*{Acknowledgment}

This work has been supported in part by Croatian Science Foundation under the Project UIP-2017-05-4968.

\bibliography{biblio} 

\begin{thebibliography}{10}

\bibitem{Kang2012HeartCA}
D.~Kang, J.~Woo, P.~J. Slomka, D.~Dey, G.~Germano, and C.-C.~J. Kuo, ``Heart
  chambers and whole heart segmentation techniques: review,'' {\em J.
  Electronic Imaging}, vol.~21, p.~010901, 2012.

\bibitem{Heimann2009StatisticalSM}
T.~Heimann and H.-P. Meinzer, ``Statistical shape models for 3d medical image
  segmentation: A review,'' {\em Medical image analysis}, vol.~13 4,
  pp.~543--63, 2009.

\bibitem{Mkel2002ARO}
T.~M{\"a}kel{\"a}, P.~Clarysse, O.~Sipil{\"a}, N.~Pauna, Q.-C. Pham, T.~Katila,
  and I.~E. Magnin, ``A review of cardiac image registration methods,'' {\em
  IEEE Transactions on Medical Imaging}, vol.~21, pp.~1011--1021, 2002.

\bibitem{Chen2020DeepLF}
C.~Chen, C.~Qin, H.~Qiu, G.~Tarroni, J.~Duan, W.~Bai, and D.~Rueckert, ``Deep
  learning for cardiac image segmentation: A review,'' {\em Frontiers in
  Cardiovascular Medicine}, vol.~7, 2020.

\bibitem{Liu2021ARO}
X.~Liu, L.~Song, S.~Liu, and Y.~Zhang, ``A review of deep-learning-based
  medical image segmentation methods,'' {\em Sustainability}, 2021.

\bibitem{Taghanaki2020DeepSS}
S.~A. Taghanaki, K.~Abhishek, J.~P. Cohen, J.~Cohen‐Adad, and G.~Hamarneh,
  ``Deep semantic segmentation of natural and medical images: a review,'' {\em
  Artificial Intelligence Review}, vol.~54, pp.~137--178, 2020.

\bibitem{Karras2018ProgressiveGO}
T.~Karras, T.~Aila, S.~Laine, and J.~Lehtinen, ``Progressive growing of gans
  for improved quality, stability, and variation,'' {\em ArXiv},
  vol.~abs/1710.10196, 2018.

\bibitem{Karras2019ASG}
T.~Karras, S.~Laine, and T.~Aila, ``A style-based generator architecture for
  generative adversarial networks,'' {\em 2019 IEEE/CVF Conference on Computer
  Vision and Pattern Recognition (CVPR)}, pp.~4396--4405, 2019.

\bibitem{Radford2016UnsupervisedRL}
A.~Radford, L.~Metz, and S.~Chintala, ``Unsupervised representation learning
  with deep convolutional generative adversarial networks,'' {\em CoRR},
  vol.~abs/1511.06434, 2016.

\bibitem{Mirza2014ConditionalGA}
M.~Mirza and S.~Osindero, ``Conditional generative adversarial nets,'' {\em
  ArXiv}, vol.~abs/1411.1784, 2014.

\bibitem{Korkinof2018HighResolutionMS}
D.~Korkinof, T.~Rijken, M.~O'Neill, J.~Yearsley, H.~Harvey, and B.~Glocker,
  ``High-resolution mammogram synthesis using progressive generative
  adversarial networks,'' {\em ArXiv}, vol.~abs/1807.03401, 2018.

\bibitem{Middel2019SynthesisOM}
L.~Middel, C.~Palm, and M.~Erdt, ``Synthesis of medical images using gans,'' in
  {\em UNSURE/CLIP@MICCAI}, 2019.

\bibitem{Zhang2018MedicalIS}
Q.~Zhang, H.~Wang, H.~Lu, D.~Won, and S.~W. Yoon, ``Medical image synthesis
  with generative adversarial networks for tissue recognition,'' {\em 2018 IEEE
  International Conference on Healthcare Informatics (ICHI)}, pp.~199--207,
  2018.

\bibitem{Shin2018MedicalIS}
H.-C. Shin, N.~A. Tenenholtz, J.~K. Rogers, C.~G. Schwarz, M.~L. Senjem, J.~L.
  Gunter, K.~P. Andriole, and M.~H. Michalski, ``Medical image synthesis for
  data augmentation and anonymization using generative adversarial networks,''
  {\em ArXiv}, vol.~abs/1807.10225, 2018.

\bibitem{Sundaram2021GANbasedDA}
S.~Sundaram and N.~Hulkund, ``Gan-based data augmentation for chest x-ray
  classification,'' {\em ArXiv}, vol.~abs/2107.02970, 2021.

\bibitem{Zhang2020MedicalIS}
H.~Zhang, Z.~Huang, and Z.~Lv, ``Medical image synthetic data augmentation
  using gan,'' {\em Proceedings of the 4th International Conference on Computer
  Science and Application Engineering}, 2020.

\bibitem{Isola2017ImagetoImageTW}
P.~Isola, J.-Y. Zhu, T.~Zhou, and A.~A. Efros, ``Image-to-image translation
  with conditional adversarial networks,'' {\em 2017 IEEE Conference on
  Computer Vision and Pattern Recognition (CVPR)}, pp.~5967--5976, 2017.

\bibitem{Abdi2019GANenhancedCE}
A.~H. Abdi, T.~S.~M. Tsang, and P.~Abolmaesumi, ``Gan-enhanced conditional
  echocardiogram generation,'' {\em ArXiv}, vol.~abs/1911.02121, 2019.

\bibitem{dataset1}
S.~Shahane, ``Mm-whs: Multi-modality whole heart segmentation.''

\end{thebibliography}
\bibliographystyle{ieeetr}

\vspace{12pt}

\end{document}